\documentclass[useAMS,usenatbib]{mn2e}
\usepackage{graphicx}

\title[WFIRST Science with a Probe Class Mission]
{WFIRST Science with a Probe Class Mission}
\author[Rodger I. Thompson, James Green, George Rieke, Xiaohui Fan, Dennis Zaristsky, Brant Robertson, Glenn Schneider, Daniel Stark Buell Jannuzi, Dennis Ebbets, Michael Kaplan and Renee Gracey]{Rodger I. Thompson$^{1}$\thanks, James Green$^{2}$ George Rieke$^{1}$, Xiaohui Fan$^{1}$, Dennis Zaristsky$^{1}$
\newauthor Brant Robertson$^{1}$, Glenn Schneider$^{1}$, Daniel Stark$^{1}$, Buell Jannuzi$^{1}$
\newauthor Dennis Ebbets$^{3}$, Michael Kaplan$^{3}$ and Renee Gracey$^{3}$\\
$^{1}$Steward Observatory, University of Arizona, Tucson, AZ 85721, USA\\
$^{2}$University of Colorado, Boulder, CO 80309, USA\\
$^{3}$Ball Aerospace \& Technology Corporation, Boulder, CO 80301}
\begin{document}

\date{Accepted xxxx. Received xxxx; in original form xxxx}

\pagerange{\pageref{firstpage}--\pageref{lastpage}} \pubyear{2011}

\maketitle

\label{firstpage}

\begin{abstract}
WFIRST is the highest priority space mission of the Decadal review,
however, it is unlikely to begin in this decade primarily due to a 
anticipated NASA budget that is unlikely to have sufficient
resources in this decade.  For this reason we present a lower
cost mission that accomplishes all of the WFIRST science as
described in the Design Reference Mission 1 with a probe
class design.   This is effort is motivated by a desire to begin
WFIRST  in a timely manner and within a budget that can fit
within the assets available to NASA on a realistic basis. The
design utilizes dichroics to form four focal planes all having
the same field of view to use the majority of available photons
from a 1.2 meter telescope.
\end{abstract}

\begin{keywords}
Instrumentation, WFIRST
\end{keywords}

\section{Introduction} \label{s-intro}

The Wide Field Infrared Survey Telescope (WFIRST) is ranked as the highest
priority space mission by the Astro 2010 Decadal Survey, New Worlds, New Horizons.
It is envisioned as a multi-purpose observatory combining wide field near infrared
imaging and spectroscopy with a microlensing survey for exoplanet detection.
In follow up studies conducted after the publishing of the Decadal Survey
telescope diameters ranging from 1.1 to 1.5 meters were considered with the
smallest diameter version performing only a subset of the Decadal Review recommended
science.  More recently the acquisition of two 2.4 meter telescope structures
referred to as the Astrophysics Focused Telescope Assets (AFTA) has motivated
the study of an expanded capability 2.4 meter version of WFIRST.  The science
capabilities of this Hubble class mission are undeniable, however, it has moved
the expected mission cost into the \$2B range and imposed significant restrictions
on the temperature of the telescope that compromises the performance at 
wavelengths longer than 2.0 microns.  An additional concern is the time scale for
implementing WFIRST.  The NASA Astrophysics Implementation Plan for 2013 
\citep{nas12}  shows the planning for a WFIRST like mission
beginning in approximately 2023, as compared to the 2013 start 
anticipated by the Decadal Survey.  Even if the late start date is taken as 
conservatively pessimistic it is appropriate to consider whether a smaller and
less expensive approach can achieve all of the expected WFIRST science.  

For this 
reason we investigate a WFIRST probe class mission where probe class in this 
context is a mission cost of $\sim$\$1B. We find that a mission with a telescope
diameter of 1.2 meters and multiple focal planes fed by a dichroic tree can 
accomplish all of the data requirements of Design Reference Mission 1 (DRM1) as
described in the WFIRST Final Report of the Science Definition Team \citet{gre12}.
We present this concept under the name WPROBE as a demonstration that the WFIRST
science goals do not need to be compromised to produce an affordable mission that
can move forward significantly sooner than the currently expected date for WFIRST.
We encourage the Science Mission Directorate at NASA headquarters to consider a
call for proposals for additional concepts that can also carry out the WFIRST 
science at significantly reduced cost and accelerated  schedule relative to the 
use of the AFTA telescope assets.

\section{DRM1 Data Requirements} \label{s-dr}
DRM1 is comprised of 4 major surveys; a High Latitude Survey (HLS) with both
imaging and low resolution spectroscopy, a supernova detection and spectroscopic
survey, a galactic plane imaging survey and a microlensing survey for exo-planet 
detection. The data requirements for these surveys are summarized in 
Table~\ref{tab-dr} taken from Table 1 of \citet{gre12}.  DRM1 also contains 
the requirement for a half year general observer program supposedly
carried out with the same filters and cameras required for the four survey programs.
\begin{table}
 \begin{minipage}{120mm}
\begin{tabular}{llll}
\hline
Survey & Filter & Area (deg)$^2$ & Depth$^a$\\  
\hline
HLS Wide Imaging & Y,J,H,K & 3400 & 26.0 \\
HLS Wide Spectroscopy & R=600 & 3400 & 1.0 x 10$^{-16}$ $^b$ \\
HLS SN 1a Imaging & J,H,K & 6.5 / 1.8 & 28.1 / 29.6 \\
HLS SN 1a Spectroscopy & R=75 & wide / deep & 27.6 / 28.5$^c$ \\
Galactic Plane & Y,J,H,K & 1240 & 25.1 \\
Microlensing & Y,W & 3.38 & n/a \\
\end{tabular}
\caption{WPROBE Characteristics}  \label{tab-dr}
$^a$Imaging 5$\sigma$; H-band AB magnitude
 
$^b$7$\sigma$ line flux in erg/cm$^2$/sec,

$^c$Continuum AB at which S/N=1 per pixel
\end{minipage}
\end{table}

\section{Instrument Concept} \label{s-ic}

Traditional imagers and imaging spectrometers utilize individual filters in sequence
to provide multicolor photometry and spectroscopy.  In this mode most of the photons
in the spectral range of the instrument are not used. More modern space instruments
such an NIRCam, being installed in JWST, utilize a dichroic to observe at two
wavelengths simultaneously.  WPROBE takes this practice to its logical next step by
utilizing a dichroic tree to observe the same field of view in 4 wavelengths
simultaneously.
\begin{figure}
  \vspace{70pt}
\resizebox{\textwidth}{!}{\includegraphics[0in,0in][14.in,3.in]{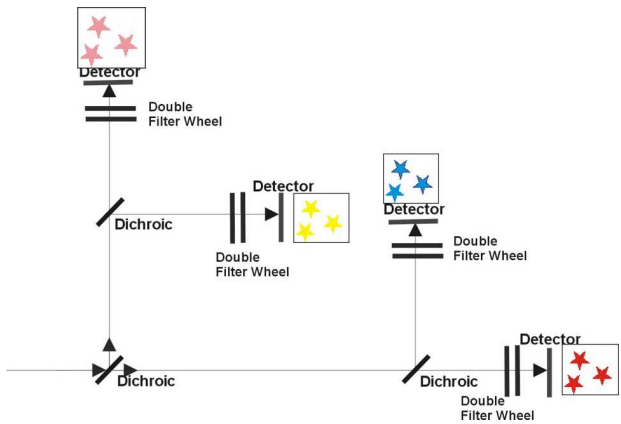}}
  \caption{The dichroic tree and focal plane concept for WPROBE} 
\label{f-dt}
\end{figure}

Modern dichroic manufacturing techniques insure very sharp transitions
from transmission to reflection to make this practice very efficient.  WPROBE has
4 focal planes to utilize the majority of photons incident on the telescope.
Fig.~\ref{f-dt} gives a schematic of the dichroic tree and focal planes.

WFIRST DRM1 observations utilize the Y, J, H and K bands with a Z band
also specified but with no DRM1 observational program associated with it.  
WPROBE utilizes dichroics to define the bands with bandwidths slightly smaller 
than the traditional filters.  The first dichroic splits the Y and J bands
from the H and K bands while the second dichroic in each arm splits
the light between the two bands. In this concept there are four focal 
planes that view the four bands simultaneously in the same field of view. 
The bandwidths are slightly smaller than
the traditional Y,J,H and K bands but the small loss in bandwidth
is more than made up for in the total speed of the system. Note that
the dichroics play the role of the filters so no additional filters
are required except for the short wavelength cutoff of the Y
filter.  The wavelength response of the detectors acts as the
long wavelength cutoff in the K band.
\begin{figure}
  \vspace{70pt}
\resizebox{\textwidth}{!}{\includegraphics[0in,0in][14.in,3.in]{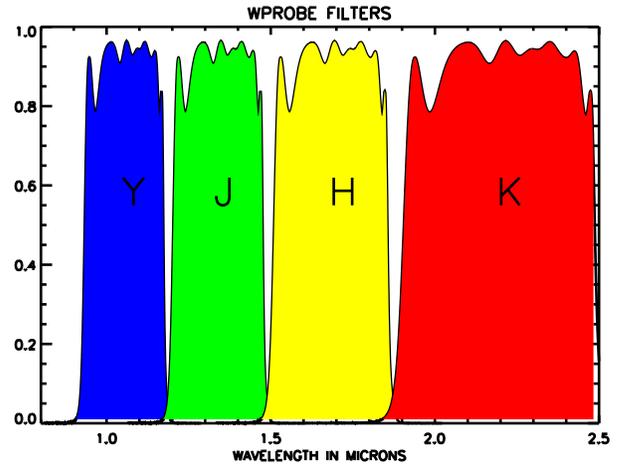}}
  \caption{A schematic of the filter bandwidths for the WPROBE Y, J, H and K filters.
The dichroic transmission and reflection properties define these bands but
for the figure the NICMOS F160W filter curves were used. The actual 
transition points and band centers are subject to further study} 
\label{fig-filt}
\end{figure}

Although the dichroics define the bands for the Y-K bands there
is the ability to define additional filters and spectroscopy.   The
dichroics are mounted on dichroic wheels that also contain a
mirror aperture and a clear aperture.  The clear apertures provide
the wide filter option for total band imaging for microlensing
studies.  The combination of clear and mirror apertures allows
complete flexibility to send light to any combination of focal
planes.  The addition of filter wheels allows more general filters
for general observer programs and provides the mechanism
for introducing the grisms required for the spectroscopic requirements
of WFIRST. Table~\ref{tab-sum} gives a summary of the basic WPROBE 
characteristics.
\begin{table}
 \begin{minipage}{120mm}
\begin{tabular}{ll}
\hline
Component & Characteristics\\  
\hline
Telescope Diameter & 1.2m  \\
Detectors & Teledyne H4RG  \\
Pixel Size & 0.18 arc sec. \\
Focal Planes & 4  \\
Dichroics & 3 \\
Detectors per Focal Plane & 3x3 \\
Science Field of View & 0.25 sq. deg. \\
Total Field of View & 0.37 sq. deg. \\
Orbit & L2 \\
Filters & Z, Y, J, H, K, Wide \\
Grisms & 2\\
Grism Resolutions & 75, 600 \\
\end{tabular}
\caption{WPROBE Characteristics preliminary estimates}  \label{tab-sum}
\end{minipage}
\end{table}

\subsection{Focal Planes} \label{ss-fp}

The WFIRST concept used for the DRM1 analysis has a pixel size of
0.18 arc seconds which we will adopt for WPROBE as well. Each of the
four focal planes contains a 3 by 3 mosaic of Teledyne H4RG 4096 by
4096 pixel detector arrays with a 2.5 micron long wavelength cutoff.
The area covered by a focal plane is 0.378 square degrees. A requirement
is to have diffraction limited image quality over at least 0.25 square degrees with
a goal of good image quality over the entire array. Preliminary
optical designs meet this requirement over the entire focal plane,
however, we will use 0.25 sq. deg. in this document since the design
is preliminary.  The one exception is for microlensing where the 
resolution requirement of 0.4 arc seconds is easily met for the
entire 0.378 sq. deg. of the focal planes.

At this point the 18 micron pixel H4RG arrays are our base line as
some problems with crosstalk have been reported for the 10 micron
pixel arrays. If these problems are eliminated by the time of the
mission we would switch to the smaller arrays.  The larger arrays 
do, however, have an advantage in reducing the amount of focal plane
area lost to gaps. Since the size of the small gap between arrays in a 
mosaic of detector arrays is independent of the size of the pixels
the percentage of gap size to array size is smaller with the larger
arrays.  The success of the Kepler mission is a significant demonstration
that mosaics much larger than the 3x3 mosaic considered here is
quite achievable.

\subsection{Spectroscopy} \label{ss-spec}

The required $\lambda / \Delta \lambda$ 75 and 600 resolution 
spectroscopy is accomplished with
grisms situated in the filter wheels between the last dichroic 
reflection or transmission and ahead of the focal planes.  Grisms have
the advantage of centering the spectrum at the location of the image of
the object in the focal plane.  This concept has been used very successfully
on both the NICMOS and WF3-IR instruments on HST.  The disadvantage of
all slitless spectrometers is that the full background flux is present
on each pixel rather than just the background in the spectral resolution
element of the pixel as with dispersing spectrometers with a slit.  The
primary background is the zodiacal light.  In all spectroscopic 
sensitivity calculations we use the same zodiacal light
model as is used in the NICMOS instrument team exposure time calculator
that successfully modeled the zodiacal flux observed by NICMOS.  The
flux has been spatially averaged since the locations of the observed
fields has not been set.  All spectral sensitivity calculations have
been done assuming all of the zodiacal light in the observed bands set
by the dichroics falls on each pixel.

\subsection{Z filter} \label{ss-zfil}

The shortest wavelength band determined by the dichroics contains both
the Y and Z bands. These bands must be isolated via filters that reside
in filter wheels as shown in Figure~\ref{f-dt}.  The Z filter is
specified in Table 1 of \citet{gre12}, however, none of the science
programs described in DRM1 use the Z filter.  Double filter wheels are
shown in front of all four focal planes in Figure~\ref{f-dt} to provide
spaces for the grisms and blanks for darks.  They are also there to
provide useful filters for a GO program and can be specified by input
from the community.  Some cost saving can be accomplished if the filter
wheels are reduced to one per focal plane and the number of filters is
set to a minimal set.

\subsection{Orbit} \label{ss-orb}

The current AFTA restrictions require operation of the telescope at a
warm temperature that adversely affects K band observations.
WPROBE, on the other hand, can exploit the advantages of an L2 orbit 
where the telescope can be cooled passivel  to a temperature that
does not compromise sensitivity in any of the bands and also provides
cooling for the detector focal planes. The passive cooling negates 
the need for cryogens or mechanical coolers resulting in significant 
cost savings over concepts that require active cooling for the
detectors or other optical components.

\section{Meeting the WFIRST Science Requirements}

Given the increased efficiency of WPROBE over traditional one filter
instruments a decreased aperture telescope is able to meet
the stated WFIRST science goals as expressed in the DRM1.  It is
very important to note that the decrease in aperture is solely to
increase the probability that WFIRST will actually happen. Of course 
a larger aperture telescope will be able to do enhanced science.
We put forward the reduced aperture WPROBE concept only to address the 
concern that the 2.4 m AFTA aperture 
requires an expenditure that puts the mission at risk.  As described
in the following sections a 1.2 m telescope is sufficient to carry out
all of the DRM1 science in a 2.6 year mission and provide another 0.4
years of General Observer science for a total 3 year mission. The only
expendable is propellant for station keeping at L2.  Enough propellant
can easily be carried for at least a 5 year mission if desired.

\subsection{High Latitude Survey Imaging Program} \label{ss-hls}

The imaging component of the High Latitude Survey (HLS) covers 3400 
square degrees to a $5\sigma$ H
band point source AB magnitude of 26 which is equivalent to $1.46 \times
10^{-7}$ janskys.  The survey covers the Y, J, H, and K band filters as 
described in Table~\ref{tab-dr}. Figure~\ref{f-hf} shows the limiting
magnitude versus time for the H band filter. The figure shows that it
takes 708 seconds to reach a SNR of 5 for an AB magnitude of 26.  With an
overhead of $10\%$ it takes 123 days to complete the survey which shows the
power of the multi-focal plane concept. The SNR of 5 limiting magnitudes in
the other bands are Y(26.8), J(26.4) and K(25.84).  To reach an AB magnitude
of 26 in all filters takes 173 days with enhanced depth in the other bands
since in the dichroic concept all bands have the same observation time.

The sensitivity calculations assume that the H4RG detectors have the same quantum 
efficiency, read noise and dark current as the NIRCam H2RG detectors for
JWST.  The physical pixel size is 18 microns rather than the smaller 10
micron pixel version.  At present time it appears that the smaller pixels
may have cross talk issues that are avoided in the larger pixels.  The
telescope and optics are radiatively cooled to 100 K and the detectors
radiatively cooled to 70 K, numbers which are easily achieved in the L2
orbit. Even though the total field of view is large enough to cover one
square degree with only three images we have done the calculations with
4 images per sq. deg.  Preliminary optical calculations indicate that there
is acceptable image quality over the total area but we have conservatively
assumed that we use only the central 0.25 sq. deg. for science where the
image quality is the best. These same characteristics are assumed in the 
other following WFIRST science programs.

\subsection{High Latitude Survey Spectroscopic Program} \label{ss-hlss}

The spectroscopic component of the HLS, also known as the Galaxy Redshift
Survey (GRS) requires a $7\sigma$ minimum detectable line flux of 
$10^{-16}$ ergs cm$^{-2}$ sec$^{-1}$ over the entire 3400 sq. deg. 
The survey is carried out with the R = 600 grisms in the wavelength range 
of 1.5 to 2.4 microns. This corresponds to the H and K bands defined by
the dichroics. The short wavelength cutoff is provided by the first dichroic
that transmits all wavelengths longer than 1.5 microns.  The second dichroic
wheel is set to the clear aperture so that the 2.5 micron cutoff of the
detector provides the long wavelength cutoff with no filters required.  

\begin{figure}
\includegraphics[width=5in,angle=90]{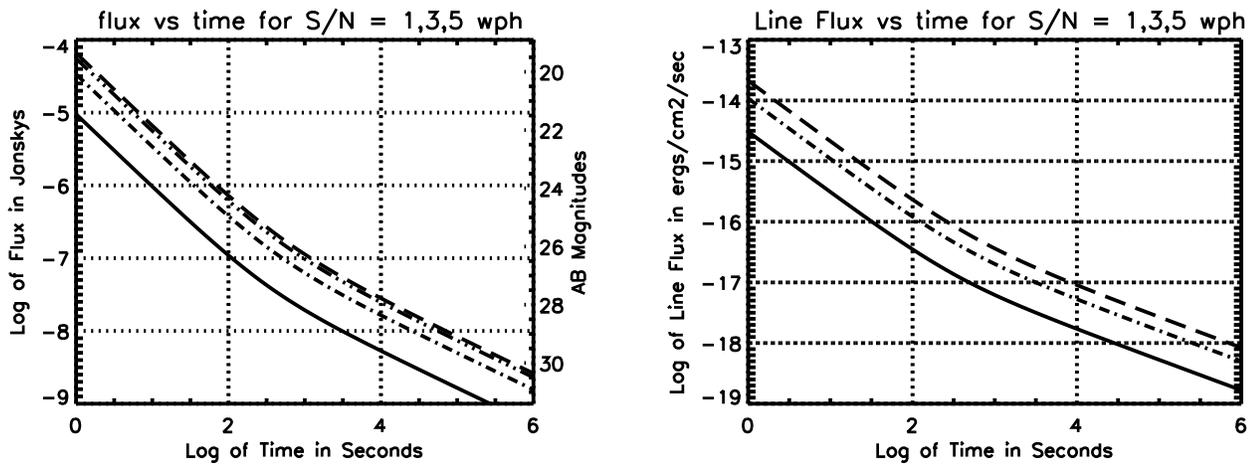}
\caption{The H band filter performance for signal to noise ratios
of 1 (lower), 3 and 5 (upper) for a single pixel.  The dash dot line
is for a 5 pixel coadded PSF at a SNR of 5.}
\label{f-hf}
\end{figure}

The right hand graph in Figure~\ref{f-hf} indicates
that it takes 526 seconds to reach that signal to noise where we have
multiplied the time to achieve $5\sigma$ by the square of 7/5.  With this
individual integration time and a $10\%$ overhead the total time needed
to accomplish the HLS spectroscopic survey is 91.1 days.

\subsection{Galactic Plane Survey} \label{ss-gps}

The Galactic Plane Survey (GPS) covers 1240 sq. deg. to a $5\sigma$ H band
point source AB magnitude of 25.1 in the Y, J, H, and K bands.  Using the
same calculations as for the HLS we find that the survey takes only 40 days
to complete, again assuming a $10\%$ overhead.

\subsection{Microlensing Exoplanet Survey} \label{ss-mes}

The Microlensing Exoplanet Survey (MES) requires a signal to noise of 100
for a J band AB magnitude of 20.5 which is $2.31 \times 10^{-5}$ janskys.
The required cadence is 15 minutes to cover a field of view greater than
2 sq. deg.  The wide filter described in DRM1 spans
0.92 to 2.5 microns.  We achieve this by rotating the first and following 
dichroic wheels to the open aperture position. This puts all of the light
onto one focal plane.  Although comments from the exoplanet community 
indicate that this is the desired option we have assumed that the wide
filter described in the DRM1 is in place to be consistent with the DRM1
program.  Under these conditions a SNR of 100 at J$_{AB}$ of 20.5 is achieved
in 3 minutes. Since the imaging requirements are quite loose, imaging
resolution of 0.4 arc seconds or less, we can use the entire focal plane
covering 0.377 sq. deg. rather than just the 0.25 sq. deg. science area.
Five exposures covers almost 2 sq. deg. satisfying the 15 minute cadence.
This can be expanded to 6 exposures to cover more than 2 sq. deg. with an
18 minute cadence or the filter can be eliminated to let all of the light
fall on the detector array to shorten the integration time. There is also
a requirement to monitor the area every 12 hours with the Y filter to 
determine color but there is no stated depth requirement. If we assume
that a $10\sigma$ measurement in the Y band is sufficient to establish 
color then only 3 seconds are needed to measure a 20.5 AB magnitude star
at that signal to noise level.  The measurement time is then set by the 
slew and settle time of the probe which is set at $<30$ seconds in the DRM1
program.

\subsection{Supernova Survey Imaging Program} \label{ss-ss}

The imaging component of the Supernova Survey (SS) has a wide and a deep
survey. The wide survey covers 6.5 sq. deg. to an AB magnitude of 28.1
and the deep survey covers 1.8 sq. deg. to an AB magnitude of 29.6 in
the J, H, and K bands.  The dichroic concept automatically includes the Y
band as well for free.  The deep band $5\sigma$ magnitude limit of 29.6
is achieved in $2.0 \times 10^5$ seconds and wide band limit is 
achieved in $1.6 \times 10^4$ seconds.  Here we assume an overhead of
100 seconds per integration rather than $10\%$ due to the longer integration
times.  We also presume that the long integration times will be broken
into several exposures with integration times sufficient to reach the
square root, background limited, signal to noise regime.  This occurs
after about 3000 seconds.  Experience with the infrared Hubble Deep Fields
has shown that this is an effective strategy.  Calculations similar to
the previous programs show that the wide survey is completed in 4.75
days and the deep survey in 16.7 days. These integration times are 
the total times, not the duration of the survey.  The temporal spacing
of the observations will be optimized for SN detection.

As an example the  total integration times for both the deep and wide 
fields is slightly less than 21.5 days. The DRM1 suggested cadence is 5
days for 0.45 years which gives about 33 visits on a 5 day cadence. 
WPROBE therefore completes the imaging survey with less than
a day of observation every 5 days for a little less than half a year.
As described in Section~\ref{ss-sss} spectroscopic observations also
occur at each imaging pointing to minimize slew times. 

\subsection{Supernova Survey Spectroscopic Program} \label{ss-sss}

The SS spectroscopic program is carried out with R=75 grism in the
0.6 to 2.0 micron spectral region.  The resolution refers to a 
2 pixel resolution element so the resolution is 150 per pixel.  The
requirement is for a signal to noise of 1 per pixel for a continuum
signal of ABmag = 27.6 for the wide (6.5 sq. deg.) and ABmag= 28.5
for the deep (1.8 sq. deg.) field. This translates to $1.04 \times
10^{-9}$ janskys for the wide field and $4.5 \times 10^{-10}$ janskys
for the deep field.  Using the H band as an example this requires
$2.5 \times 10^5$ seconds for the wide field and $1.3 \times 10^6$
seconds for the deep field.  These are cumulative times not individual
exposure times.  The total time required is 75.5 days for the wide 
field and 110 days for the deep field using only the 0.25 sq. deg.
science area of the focal plane.  Individual exposure times will be
on the order of an hour so the overhead times in this program are
minimal.  The DRM1 reserves 1.8 years for this program but does not
state a cadence. We assume that on each visit for the imaging program
two days are spent on the spectroscopic program. At each pointing
for the imaging program spectroscopic observations are also taken
for a longer period of time. After the imaging program has been completed
the spectroscopic program will continue in the remaining 1.4 years
allotted to it.

\section{Total Program Time for the WFIRST DRM1 Science}

The times expressed in Table~\ref{tab-pt} are the total integration
times needed to reach the sensitivity limits expressed in
DRM1.  The individual integrations are arranged to optimize
efficiency and to best sample temporal events.
\begin{table}
 \begin{minipage}{120mm}
\begin{tabular}{ll}
\hline
Program & Time in Days\\  
\hline
HLS Imaging & 173  \\
HLS Spectroscopic (GRS) & 91.1  \\
Galactic Plane Imaging & 40 \\
Microlensing Survey & 438$^a$  \\
Supernova Imaging Wide & 4.75 \\
Supernova Imaging Deep & 16.7 \\
Supernova Spectroscopy Wide & 75.5 \\
Supernova Spectroscopy Deep & 110 \\
Total Program & 949 (2.6 years) \\
\hline
\end{tabular}
\caption{WPROBE Program Times}  \label{tab-pt}
$^a$Set by requested duration
\end{minipage}
\end{table}

\section{WPROBE Relative to WNEW}

Although the WPROBE dichroic concept presented here is oriented
toward a low cost probe mission that can carry out all of the
WFIRST DRM1 science it can also be used to greatly enhance
the efficacy of an AFTA mission. WPROBE is similar in concept 
to the WNEW mission presented at the SALSO conference in Huntsville
Alabama in February of 2013 on the use of AFTA for other purposes. 
It is clear that a 2.4m version of WPROBE or WNEW will accomplish more 
science but at a significantly larger cost.

\clearpage
\label{lastpage}
\end{document}